\documentclass[pdflatex,sn-mathphys-num,iicol]{sn-jnl}

\usepackage{graphicx}
\makeatletter
\providecommand\@reinserts{}
\makeatother
\usepackage{manyfoot}
\usepackage{amsmath,amssymb,amsfonts,mathtools,bm}
\usepackage{mathrsfs}
\usepackage{xcolor}
\usepackage{hyperref}
\usepackage{microtype}

\newcommand{\dd}{\mathrm d}

\newcommand{\PA}{\mathcal P_A}
\newcommand{\PB}{\mathcal P_B}
\newcommand{\TT}{\mathrm{TT}}
\newcommand{\Mpl}{M_{\rm Pl}}
\newcommand{\br}{\mathrm{br}}
\newcommand{\proj}{\mathrm{proj}}
\newcommand{\eff}{\mathrm{eff}}

\newcommand{\scri}{\mathscr I}

\begin{document}

\title[Radiative matching across a projector bridge]{Radiative Matching Across a Projector Bridge}

\author[1]{\fnm{M. E.} \sur{Abishev}}
\author[1]{\fnm{D. Z.} \sur{Berkimbayev}}

\affil[1]{Al-Farabi Kazakh National University, Al-Farabi av. 71, Almaty 050040, Kazakhstan}

\abstract{Two distinct Einstein--Hilbert sectors are related by a covariant correspondence bridge that maps stress-energy through bitensor kernels. The Bianchi identities enforce conservation, with bridge stresses balancing cross-sector exchange. For positive gravitational couplings and paired Minkowski backgrounds, a fixed metric-independent bridge leaves the quadratic action block diagonal, so the vacuum contains two massless spin-2 fields and no Fierz--Pauli mass term. A reciprocal first-order transport law generates a flux-preserving rotation of the canonical outgoing data. This divides the emitted power between visible and hidden channels while preserving the total quadrupole luminosity and the general-relativistic chirp phase. The reduced visible amplitude produces a systematic standard-siren distance shift, whose present sensitivity can be estimated from gravitational-wave catalog cosmology combined with an independent distance calibration. A source-level rescaling is physically inequivalent: it changes radiation reaction and requires compensating bridge energy flow. The construction therefore describes amplitude transfer between massless tensor channels, distinct from massive-graviton dispersion and propagation-induced oscillations.}

\keywords{gravitational waves, standard sirens, radiative matching, hidden sectors, bimetric gravity, effective gravity}

\maketitle
\begingroup
\renewcommand{\thefootnote}{}
\footnotetext{D. Z. Berkimbayev: \href{mailto:daulet9432@gmail.com}{daulet9432@gmail.com}}
\endgroup

\section{Introduction}\label{sec:introduction}

Many two-metric theories place both metrics on one differentiable manifold and couple them through a nonderivative potential. The Fierz--Pauli term is the linear prototype, while ghost-free nonlinear bimetric gravity requires a special interaction potential to remove the Boulware--Deser mode~\cite{FierzPauli1939,BoulwareDeser1972,deRham2011,HassanRosen2012,Hinterbichler2012,deRham2014,SchmidtMay2016}. Their tensor eigenmodes can acquire different propagation phases and produce graviton oscillations~\cite{DeFelice2014,Brizuela2024}. Kernel-based gravity and bimetric modified Newtonian dynamics provide different routes to effective cross-sector sources~\cite{HehlMashhoon2009,Milgrom2009}.

The construction considered here instead starts from two distinct Lorentzian spacetimes,
\begin{equation}
 (M_A,g^A_{\mu\nu}),\qquad (M_B,g^B_{ab}),
 \label{eq:two-spacetimes}
\end{equation}
with independent Einstein--Hilbert dynamics~\cite{Einstein1916,Hilbert1915}. In four dimensions these are the standard metric sectors selected by Lovelock's theorem under the usual assumptions of locality, diffeomorphism covariance, and second-order field equations~\cite{Lovelock1971}. Here, we do not introduce algebraic interaction $V(g_A^{-1}\phi^*g_B)$, but instead, the bridge projects energy-momentum data between the spacetimes. This field-equation construction is used as a low-energy effective description~\cite{Donoghue1994,Burgess2004} and implements the two-sector gravitational setting motivated in Ref.~\cite{AbishevBerkimbayev2026}. Standard conventions of general relativity (GR) are used throughout~\cite{Wald1984,Carroll2004}.

The main question is where the bridge acts on gravitational radiation. Acting on the outgoing radiative data preserves the GR near-zone generation and divides the emitted flux between two massless channels. Acting on the local source changes radiation reaction and introduces an additional bridge energy current. We construct the conserved projected energy-momentum tensors, determine the fixed-bridge vacuum spectrum, and derive the radiative power and standard-siren relations from a minimal two-channel transport law. The weak-field and homogeneous limits are retained in Appendix~\ref{app:nonradiative}.

\section{Correspondence bridge with projected energy-momentum}\label{sec:bridge}

Because the manifolds are distinct, a cross-sector source map must specify both a point correspondence and a tangent-space map. Greek indices refer to $M_A$ manifold and Latin indices to $M_B$. We introduce a correspondence space
\begin{equation}
 C\subset M_A\times M_B,
 \label{eq:correspondence-space}
\end{equation}
with projections $\pi_A:C\to M_A$ and $\pi_B:C\to M_B$. In the local single-valued limit, $C$ is the graph of a map $\phi:M_A\to M_B$, so that $y=\phi(x)$.

At a paired point $(x,y)\in C$, the bridge carries maps
\begin{align}
 J^a{}_{\mu}(x,y)&:T_xM_A\longrightarrow T_yM_B,\nonumber\\
 K^\mu{}_{a}(x,y)&:T_yM_B\longrightarrow T_xM_A.
 \label{eq:J-K-maps}
\end{align}
They form a reflexive generalized-inverse pair on a nondegenerate coupled subspace,
\begin{equation}
 JKJ=J,\qquad KJK=K,
 \label{eq:generalized-inverse}
\end{equation}
and are metric-adjoint there, $K=J^\dagger$. The induced tangent-space projectors are
\begin{equation}
 (P_A)^\mu{}_{\nu}=K^\mu{}_{a}J^a{}_{\nu},\qquad
 (P_B)^a{}_{b}=J^a{}_{\mu}K^\mu{}_{b},
 \label{eq:tangent-projectors}
\end{equation}
with $P_A^2=P_A$ and $P_B^2=P_B$.

A covariant source map from $M_B$ to $M_A$ is
\begin{align}
 \PA[T_B]_{\mu\nu}(x)
 ={}&\int_{M_B}W_A(x,y)J^a{}_{\mu}(x,y)J^b{}_{\nu}(x,y)
 \nonumber\\
 &\times T^B_{ab}(y)\,\dd\mu_B(y),
 \label{eq:kernel-A}
\end{align}
where $\dd\mu_B=\sqrt{-g_B}\,\dd^4y$. The reverse map is
\begin{align}
 \PB[T_A]_{ab}(y)
 ={}&\int_{M_A}W_B(y,x)K^\mu{}_{a}(x,y)K^\nu{}_{b}(x,y)
 \nonumber\\
 &\times T^A_{\mu\nu}(x)\,\dd\mu_A(x).
 \label{eq:kernel-B}
\end{align}
For dimensionless $J$ and $K$, the weights have length dimension $-4$. The local limit is obtained with a covariant delta density, for example
\begin{equation}
 W_A(x,y)\longrightarrow\delta_B[y,\phi(x)],
\end{equation}
which gives $\PA[T_B]_{\mu\nu}=J^a{}_{\mu}J^b{}_{\nu}T^B_{ab}$ at $y=\phi(x)$. Pointwise idempotency of Eq.~\eqref{eq:tangent-projectors} does not imply idempotency of a genuinely nonlocal integral operator.

The gravitational action is
\begin{align}
 S_{\rm grav}={}&\frac{1}{16\pi G_A}\int_{M_A}\dd^4x
 \sqrt{-g_A}(R_A-2\Lambda_A)\nonumber\\
 &+\frac{1}{16\pi G_B}\int_{M_B}\dd^4y
 \sqrt{-g_B}(R_B-2\Lambda_B).
 \label{eq:two-EH-action}
\end{align}
No bridge-mediated algebraic interaction constructed from a pullback of $g_B$ to $M_A$, such as $V(g_A^{-1}\phi^*g_B)$ in the local correspondence limit, is included. At the effective field-equation level,
\begin{align}
 G^A_{\mu\nu}+\Lambda_Ag^A_{\mu\nu}&=8\pi G_A\Theta^A_{\mu\nu},\label{eq:field-A}\\
 G^B_{ab}+\Lambda_Bg^B_{ab}&=8\pi G_B\Theta^B_{ab},\label{eq:field-B}
\end{align}
where
\begin{align}
 \Theta^A_{\mu\nu}&=T^A_{\mu\nu}+\epsilon_A\PA[T_B]_{\mu\nu}+T^{A,\br}_{\mu\nu},\label{eq:Theta-A}\\
 \Theta^B_{ab}&=T^B_{ab}+\epsilon_B\PB[T_A]_{ab}+T^{B,\br}_{ab}.\label{eq:Theta-B}
\end{align}
The Bianchi identities require
\begin{equation}
 \nabla_A^\mu\Theta^A_{\mu\nu}=0,\qquad
 \nabla_B^a\Theta^B_{ab}=0.
 \label{eq:Bianchi}
\end{equation}
If the projected sources are not separately conserved, the divergences of the bridge stresses must compensate the corresponding exchange currents. Equation~\eqref{eq:Bianchi} is a necessary consistency condition.

The static and homogeneous limits of the same source tensor are collected in Appendix~\ref{app:nonradiative}.

We now take $G_A>0$ and $G_B>0$. The cosmological constants and zeroth-order effective sources are chosen so that the paired Minkowski metrics solve Eqs.~\eqref{eq:field-A} and~\eqref{eq:field-B}. For fixed metric-independent bridge data, the quadratic metric action is then block diagonal,
\begin{align}
 S_2={}&\frac{1}{64\pi G_A}\int\dd^4x\,
 h_A^{\mu\nu}{\cal E}_{\mu\nu}{}^{\alpha\beta}h^A_{\alpha\beta}\nonumber\\
 &+\frac{1}{64\pi G_B}\int\dd^4y\,
 h_B^{ab}{\cal E}_{ab}{}^{cd}h^B_{cd}+S_{\rm src}.
 \label{eq:quadratic-action}
\end{align}
No Fierz--Pauli structure proportional to $(h^A_{\mu\nu}-h^B_{\mu\nu})^2-(h_A-h_B)^2$ occurs. In de Donder gauge, the vacuum equations are
\begin{equation}
 \Box_A\bar h^A_{\mu\nu}=0,\qquad
 \Box_B\bar h^B_{ab}=0,
 \label{eq:vacuum-waves}
\end{equation}
with $k_A^2=k_B^2=0$. The fixed, metric-independent bridge therefore leaves two massless vacuum spin-2 fields.

The result can be compared directly with a bimetric mass interaction. On a common manifold such an interaction has the form
\begin{equation}
 S_{\rm int}=m^2\Mpl^2\int\dd^4x\sqrt{-g_A}\,V(g_A^{-1}g_B),
\end{equation}
whose Fierz--Pauli expansion contains
\begin{align}
 S_{\rm mass}=-\frac{m^2\Mpl^2}{8}\int\dd^4x
 &\big[(h^A_{\mu\nu}\nonumber \\-h^B_{\mu\nu})^2&-(h_A-h_B)^2\big].
 \label{eq:Fierz-Pauli-comparison}
\end{align}
Equation~\eqref{eq:quadratic-action} contains no such term. The correspondence bridge therefore changes the projected sources without shifting either vacuum propagator away from its massless pole.

\section{Asymptotic flux-isometric radiative matching}\label{sec:asymptotic}

The radiative benchmark is defined after the near-zone waveform has been generated. For asymptotically flat sectors, let $\scri_A^+$ and $\scri_B^+$ denote future null infinity. Near each component of null infinity we use Bondi-like coordinates $(u_i,r_i,\Omega_i)$. A matched asymptotic identification $\varphi_\infty:\scri_A^+\to\scri_B^+$ is assumed to identify the normalized retarded times, angular coordinates, polarization bases, and asymptotic time-translation generators of the two sectors.

The matching should be understood as an effective condition on outgoing asymptotic states, analogous to an $S$-matrix boundary condition. When a strict null infinity is unavailable, the same construction can be imposed on matched large-radius wave-zone worldtubes and then taken to the asymptotic limit. At finite radius the leading luminosity is
\begin{align}
 P_i(r_i,u_i)=\frac{r_i^2}{32\pi G_i}
 \int \dd\Omega_i\,
 \partial_{u_i}h^{\TT}_{i,AB}\,
 \partial_{u_i}h_i^{\TT,AB}\nonumber \\+{\cal O}(r_i^{-1}).
 \label{eq:finite-radius-power}
\end{align}

The detector strain falls as $r_i^{-1}$. The finite radiative coefficient at $\scri_i^+$ is therefore
\begin{equation}
 C_{i,AB}(u_i,\Omega_i)
 \equiv\lim_{r_i\to\infty}r_i h^{\TT}_{i,AB}(u_i,r_i,\Omega_i),
 \label{eq:asymptotic-shear}
\end{equation}
with news tensor
\begin{equation}
 N_{i,AB}\equiv\partial_{u_i}C_{i,AB}.
 \label{eq:news-tensor}
\end{equation}
In Bondi language, $C_{i,AB}$ is the radiative shear and $N_{i,AB}$ is the news~\cite{Bondi1962,Sachs1962}. The transverse--traceless decomposition is performed only after an asymptotic background and boundary conditions have been fixed.

We introduce canonically normalized radiative data
\begin{equation}
 \psi_{i,AB}\equiv\frac{C_{i,AB}}{\sqrt{G_i}}.
 \label{eq:canonical-radiative-field}
\end{equation}
The radiative phase space is understood modulo configurations with vanishing news. At each retarded time its angular--polarization flux form is
\begin{equation}
 (\dot\psi_1,\dot\psi_2)_{u,i}
 =\frac{1}{32\pi}\int\dd\Omega_i\,
 \partial_{u_i}\psi_{1,AB}\,
 \partial_{u_i}\psi_2^{AB},
 \label{eq:instantaneous-flux-form}
\end{equation}
so that
\begin{equation}
 P_i(u_i)=(\dot\psi_i,\dot\psi_i)_{u,i},
 \qquad
 E_i^{\rm rad}=\int\dd u_i\,P_i(u_i).
 \label{eq:power-and-energy}
\end{equation}
Equivalently, the integrated flux bilinear is
\begin{equation}
 \langle\psi_1,\psi_2\rangle_{F,i}
 =\int\dd u_i\,(\dot\psi_1,\dot\psi_2)_{u,i}.
 \label{eq:flux-inner-product}
\end{equation}
Up to the standard averaging convention, this is the gravitational-wave energy flux norm~\cite{Maggiore2008}.

The asymptotic tangent map induced by $J$ acts on symmetric tensors and is followed by the asymptotic TT projection. Schematically,
\begin{equation}
 U_{\rm rad}={\cal N}\,\Pi_B^{\TT}\circ(J_\infty\otimes J_\infty)\circ\Pi_A^{\TT},
 \label{eq:U-from-J}
\end{equation}
where ${\cal N}$ fixes the flux normalization. The benchmark assumes that $U_{\rm rad}$ is independent of the matched retarded time and is unitary on the participating angular--polarization radiative subspaces at each $u$,
\begin{equation}
 (U_{\rm rad}\dot\psi_1,U_{\rm rad}\dot\psi_2)_{u,B}
 =(\dot\psi_1,\dot\psi_2)_{u,A}.
 \label{eq:U-unitary}
\end{equation}
Consequently it also preserves the integrated form in Eq.~\eqref{eq:flux-inner-product}. The matching acts only on the gauge-invariant outgoing radiative data after the local constraints have been solved.

Let $U_{\rm src}$ denote the corresponding map between the participating TT source-data spaces after the background and TT decomposition have been fixed. If ${\cal T}_i$ is the retarded transfer operator from those source data to outgoing canonical radiation, consistency with a radiative map induced by the correspondence bridge requires
\begin{equation}
 {\cal T}_B U_{\rm src}=U_{\rm rad}{\cal T}_A.
 \label{eq:intertwining}
\end{equation}
Equation~\eqref{eq:intertwining} summarizes the matched-propagation assumption.

To obtain the channel map from a transport law, introduce a dimensionless bridge parameter $s\in[0,1]$ and collect the participating radiative data into
\begin{equation}
 \boldsymbol{\Psi}(u,s)=
 \begin{pmatrix}
  \psi_A(u,s)\\[2pt]
  \psi_B(u,s)
 \end{pmatrix}.
 \label{eq:transport-state}
\end{equation}
The parameter $s$ labels the matched transfer rather than an additional spacetime direction.
The minimal reciprocal first-order coupling consistent with $U_{\rm rad}$ is
\begin{align}
 \partial_s\boldsymbol{\Psi}
 &=\kappa(s)\,\mathbb J\boldsymbol{\Psi},
 \label{eq:transport-law}\\
 \mathbb J&=
 \begin{pmatrix}
  0&-U_{\rm rad}^{\dagger}\\
  U_{\rm rad}&0
 \end{pmatrix}.
 \label{eq:transport-generator}
\end{align}
Here $\kappa(s)$ is real and independent of retarded time. Unitarity of $U_{\rm rad}$ gives $\mathbb J^{\dagger}=-\mathbb J$ and $\mathbb J^2=-I$, and therefore
\begin{equation}
 \frac{\dd}{\dd s}
 \left\lVert\partial_u\boldsymbol{\Psi}\right\rVert_F^2=0.
 \label{eq:transport-flux-conservation}
\end{equation}
The transport law thus preserves the direct-sum flux norm. Since its generator has a fixed channel direction, its solution depends only on
\begin{equation}
 \theta=\int_0^1\kappa(s)\,\dd s.
 \label{eq:integrated-mixing-angle}
\end{equation}
Using $\mathbb J^2=-I$, the finite transport operator is
\begin{align}
 {\mathfrak R}_\theta=&e^{\theta\mathbb J}=
 \begin{pmatrix}
 \cos\theta\,I_A & -\sin\theta\,U_{\rm rad}^\dagger\\
 \sin\theta\,U_{\rm rad} & \cos\theta\,I_B
 \end{pmatrix},
 \\ &0\leq\theta\leq\frac{\pi}{2}.
 \label{eq:asymptotic-rotation}
\end{align}
Here $I_A$ and $I_B$ are the identities on the participating radiative subspaces. On the complete radiative phase space the operator is extended as
\begin{equation}
 {\mathfrak R}^{\rm full}_\theta
 ={\mathfrak R}_\theta\oplus I_{A,\perp}\oplus I_{B,\perp}.
 \label{eq:full-rotation}
\end{equation}
The bridge geometry fixes the channel map $U_{\rm rad}$, while the accumulated coupling in Eq.~\eqref{eq:integrated-mixing-angle} fixes the single angle of the conservative two-channel transfer.

For a GR waveform generated in sector $A$ with no incoming sector-$B$ radiation,
\begin{equation}
 \begin{pmatrix}
 \psi_A^{\rm out}\\[2pt]
 \psi_B^{\rm out}
 \end{pmatrix}
 ={\mathfrak R}_\theta
 \begin{pmatrix}
 \psi_{\rm GR}^{\rm out}\\[2pt]
 0
 \end{pmatrix},
 \label{eq:outgoing-matching}
\end{equation}
so that
\begin{equation}
 \psi_A^{\rm out}=\cos\theta\,\psi_{\rm GR}^{\rm out},\qquad
 \psi_B^{\rm out}=\sin\theta\,U_{\rm rad}\psi_{\rm GR}^{\rm out}.
 \label{eq:canonical-output}
\end{equation}
The asymptotic shear coefficients satisfy
\begin{equation}
 C_A^{\rm out}=\cos\theta\,C_{\rm GR}^{\rm out},\quad
 C_B^{\rm out}=\sqrt{\frac{G_B}{G_A}}\sin\theta\,
 U_{\rm rad}C_{\rm GR}^{\rm out}.
 \label{eq:shear-output}
\end{equation}
Detector strains at matched luminosity distances inherit the same amplitude factors. For matched waveform shapes,
\begin{equation}
 \frac{\|h_B\|}{\|h_A\|}
 =\sqrt{\frac{G_B}{G_A}}\tan\theta.
 \label{eq:amplitude-ratio}
\end{equation}

\begin{figure}[t]
\centering
\includegraphics[width=0.94\linewidth]{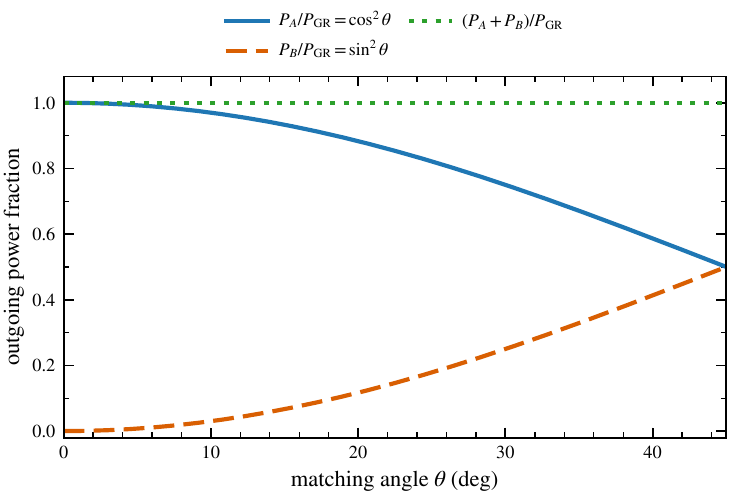}
\caption{Outgoing power fractions in the flux-isometric matching ansatz. The displayed interval corresponds to a sector-$A$-dominated signal; the full physical range is $0\leq\theta\leq\pi/2$}
\label{fig:power-partition}
\end{figure}

\begin{figure}[t]
\centering
\includegraphics[width=0.94\linewidth]{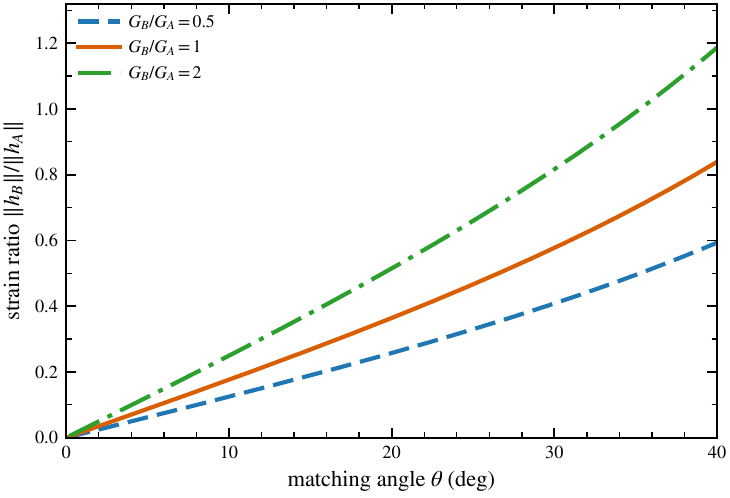}
\caption{Hidden-to-visible strain ratio for matched waveform shapes. The coupling ratio affects detector-level strain amplitudes, while the canonical flux split remains $\cos^2\theta$ and $\sin^2\theta$}
\label{fig:amplitude-ratio}
\end{figure}

Because Eq.~\eqref{eq:U-unitary} holds at each matched retarded time, flux unitarity gives
\begin{align}
 P_A(u)=\cos^2\theta\,P_{\rm GR}(u),\\
 P_B(u)=\sin^2\theta\,P_{\rm GR}(u),
 \label{eq:power-partition}
\end{align}
and therefore
\begin{equation}
 P_A(u)+P_B(u)=P_{\rm GR}(u).
 \label{eq:power-conservation}
\end{equation}
Integration over retarded time yields
\begin{equation}
 E_A^{\rm rad}+E_B^{\rm rad}=E_{\rm GR}^{\rm rad}.
 \label{eq:energy-conservation}
\end{equation}
The corresponding hidden power fraction is
\begin{equation}
 \eta_B\equiv\frac{P_B(u)}{P_{\rm GR}(u)}=\sin^2\theta,
 \label{eq:eta-theta}
\end{equation}
with the same ratio for the integrated radiated energies. An observer confined to sector $A$ measures $\cos^2\theta$ and is insensitive to the sign of $\theta$. Figures~\ref{fig:power-partition} and~\ref{fig:amplitude-ratio} focus on $0\leq\theta\leq\pi/4$, where sector $A$ remains the dominant outgoing channel; the physical range of the ansatz is the full interval in Eq.~\eqref{eq:asymptotic-rotation}.

The binary generates the ordinary GR outgoing waveform and loses energy at the GR quadrupole rate,
\begin{equation}
 \dot E_{\rm orb}=-P_{\rm GR}.
 \label{eq:GR-near-zone-balance}
\end{equation}
The asymptotic rotation redistributes this flux according to Eq.~\eqref{eq:power-partition}. Consequently the leading chirp phase retains its GR form while the visible amplitude is multiplied by $\cos\theta$.

Consider the local prescription in which the sector-$A$ wave equation is sourced by $\cos\theta\,T_A^{\TT}$. At Newtonian order the Burke--Thorne reaction potential is linear in the effective quadrupole~\cite{PoissonWill2014,Burke1971},
\begin{equation}
 \Phi_{\rm react}=-\frac{G_A}{5}\,x^ix^j
 \frac{\dd^5\widetilde Q_{ij}}{\dd t^5},
 \qquad \widetilde Q_{ij}=\cos\theta\,Q_{ij},
 \label{eq:Burke-Thorne}
\end{equation}
where $c=1$. If the force acts on the unscaled matter quadrupole, the orbital work is
\begin{equation}
 \dot E_{\rm orb}=-\cos\theta\,P_{\rm GR}.
 \label{eq:source-level-balance}
\end{equation}
The radiative amplitude produced by the same source is proportional to $\cos\theta$, and its sector-$A$ flux is therefore
\begin{equation}
 P_A^{\rm rad}=\cos^2\theta\,P_{\rm GR}.
 \label{eq:source-level-flux}
\end{equation}
Equations~\eqref{eq:source-level-balance} and~\eqref{eq:source-level-flux} do not close the energy balance by themselves. For this single-channel source prescription, conservation requires the bridge current
\begin{equation}
 P_{\rm br}= -\dot E_{\rm orb}-P_A^{\rm rad}
 =\cos\theta(1-\cos\theta)P_{\rm GR}.
 \label{eq:bridge-energy-current}
\end{equation}
If additional radiative channels participate, their fluxes enter the same balance equation. This is the radiative counterpart of the exchange stresses already required by Eq.~\eqref{eq:Bianchi}.

The leading frequency evolution of the source-rescaled prescription obeys
\begin{equation}
 \dot\omega\propto\cos\theta\,{\cal M}_c^{5/3}\omega^{11/3},
\end{equation}
so that
\begin{equation}
 \left({\cal M}_c^{\eff}\right)^{5/3}
 =\cos\theta\,{\cal M}_c^{5/3}.
 \label{eq:effective-chirp-mass}
\end{equation}
The visible leading-order strain becomes
\begin{equation}
 h_A\propto\frac{\cos\theta\,{\cal M}_c^{5/3}}{D_L}
 =\frac{({\cal M}_c^{\eff})^{5/3}}{D_L}.
\end{equation}
Thus the amplitude factor is absorbed into the chirp mass in this specific source prescription and the leading distance residual vanishes. The asymptotic rotation instead preserves Eq.~\eqref{eq:GR-near-zone-balance}, keeps the GR chirp mass, and produces the distance shift derived below.

This construction also differs sharply from massive and ghost-free bimetric gravity. Those theories contain a massive spin-2 pole or a massive tensor eigenmode whose propagation phase can generate dispersion and graviton oscillations~\cite{DeFelice2014,Brizuela2024}. Equations~\eqref{eq:vacuum-waves} contain two massless poles. The rotation~\eqref{eq:asymptotic-rotation} therefore transfers amplitude between outgoing channels without producing a massive-dispersion phase or propagation-induced oscillations.

\section{Interpretation}\label{sec:sirens}

\begin{figure}[t]
\centering
\includegraphics[width=0.98\linewidth]{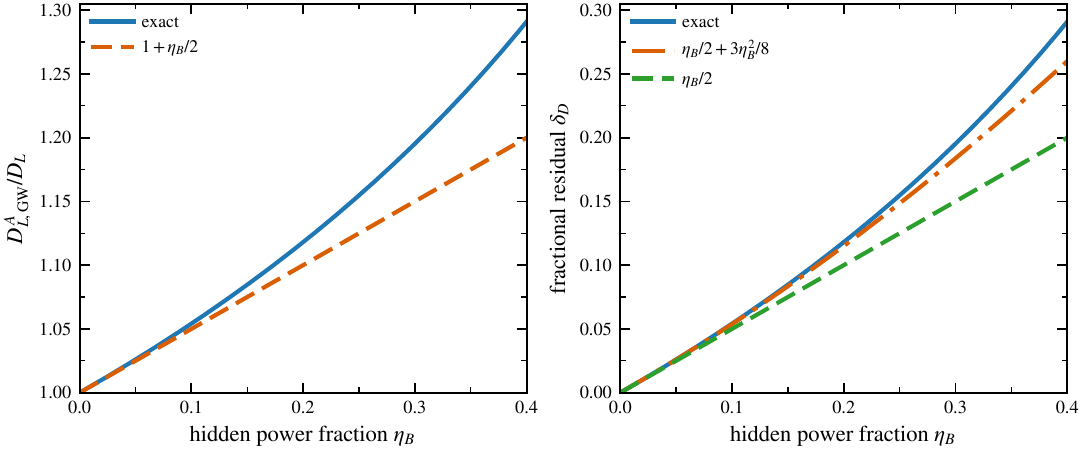}
\caption{Standard-siren observables of the asymptotic matching. Left: exact gravitational-wave distance ratio and its leading expansion. Right: exact fractional residual with its linear and quadratic weak-leakage expansions}
\label{fig:siren-observables}
\end{figure}

A detector confined to sector $A$ measures the reduced strain $h_A=\cos\theta\,h_{\rm GR}$. Under a GR waveform analysis, this is interpreted as a larger gravitational-wave luminosity distance,
\begin{equation}
 D^A_{L,{\rm GW}}=D_L\sec\theta
 =\frac{D_L}{\sqrt{1-\eta_B}}.
 \label{eq:distance-bias}
\end{equation}
The fractional residual is
\begin{equation}
 \delta_D=(1-\eta_B)^{-1/2}-1
 =\frac{\eta_B}{2}+\frac{3\eta_B^2}{8}+{\cal O}(\eta_B^3).
 \label{eq:distance-residual}
\end{equation}
The exact distance ratio and its weak-leakage approximations are displayed together in Fig.~\ref{fig:siren-observables}.

A constant factor is degenerate with the absolute distance scale. At low redshift,
\begin{equation}
 H_{0,\rm inf}=H_{0,\rm true}\cos\theta
 =H_{0,\rm true}\sqrt{1-\eta_B}.
 \label{eq:H0-degeneracy}
\end{equation}
An electromagnetic counterpart supplies a host redshift and can improve the inclination measurement. The degeneracy is broken only when bright sirens are combined with an independent absolute-distance or cosmological calibration, or when the bridge factor depends on redshift, frequency, direction, or source properties.

For an independently calibrated value $H_{0,\rm cal}$, Eq.~\eqref{eq:H0-degeneracy} gives the direct estimator
\begin{equation}
 \widehat q\equiv\frac{H_{0,\rm GW}}{H_{0,\rm cal}}
 =\cos\theta,\qquad
 \eta_B=1-\widehat q^{\,2},
 \label{eq:catalog-estimator}
\end{equation}
with $0<\widehat q\leq1$ for the attenuation considered here and independent fractional uncertainty
\begin{equation}
 \left(\frac{\sigma_q}{\widehat q}\right)^2
 \simeq
 \left(\frac{\sigma_{\rm GW}}{H_{0,\rm GW}}\right)^2
 +\left(\frac{\sigma_{\rm cal}}{H_{0,\rm cal}}\right)^2.
 \label{eq:q-uncertainty}
\end{equation}
The GWTC-5.0 cosmology analysis~\cite{AbacCosmologyGWTC5}, extending the catalog method of Ref.~\cite{AbbottCosmologyGWTC3}, reports $H_{0,\rm GW}=71.0^{+9.0}_{-7.1}\,\mathrm{km\,s^{-1}\,Mpc^{-1}}$. If the calibration uncertainty is subdominant and the central values agree with $\widehat q=1$, the present fractional precision gives $\sigma_q\simeq0.113$ and the one-sided $95\%$ sensitivity
\begin{equation}
 \eta_B\lesssim
 1-\left(1-1.645\,\sigma_q\right)^2
 \simeq0.34.
 \label{eq:current-sensitivity}
\end{equation}
This estimate sets the current scale of the test; a catalog constraint follows by including $\widehat q$ in the standard population and calibration analysis.

The standard-siren method was introduced in
Refs.~\cite{Schutz1986,HolzHughes2005}, and GW170817 established its
multimessenger implementation~\cite{AbbottGW170817,AbbottStandardSiren2017}.
Current gravitational-wave tests find no evidence for massive-graviton
dispersion~\cite{AbbottGWTC3}, consistently with the two massless tensor
channels of Eq.~\eqref{eq:vacuum-waves}. These tests primarily constrain
frequency- or redshift-dependent propagation effects~\cite{Belgacem2018} and therefore do not
directly measure the constant amplitude transfer predicted here. The bridge
parameter is instead tested through a common offset between gravitational-wave
luminosity distances and an independently calibrated distance scale.

\section{Conclusions}\label{sec:conclusion}

We have constructed a covariant two-spacetime bridge at the level of projected energy-momentum tensors. The generalized-inverse tangent maps define the participating subspaces, the bitensor kernels transport the sources, and the bridge stresses enforce the two Bianchi identities. The Newtonian and matched-FLRW limits follow from the same source tensor and are summarized in Appendix~\ref{app:nonradiative}.

For positive gravitational couplings and paired Minkowski backgrounds, a fixed metric-independent bridge leaves the quadratic metric action block diagonal. The spectrum therefore consists of two massless spin-2 fields. This distinguishes the construction from massive and ghost-free bimetric gravity before any radiative observable is considered.

The radiative bridge is the finite solution of the reciprocal transport law~\eqref{eq:transport-law}, with $\theta$ fixed by the integrated coupling~\eqref{eq:integrated-mixing-angle}. Its consequences are fixed by the flux norm. The total quadrupole luminosity and GR chirp phase are preserved, while the visible amplitude is reduced. A source-rescaled prescription is not equivalent: it changes radiation reaction and requires the bridge current in Eq.~\eqref{eq:bridge-energy-current}. This establishes the physical distinction between local source projection and asymptotic radiative transfer.

The bridge geometry determines the allowed source and radiative maps, while the integrated coupling determines the single angle of their conservative two-channel realization. Within this realization, the power partition and standard-siren relation are exact and provide the characteristic gravitational-wave signature of the projector bridge. Current catalog precision corresponds to sensitivity to hidden power fractions of order a few tenths when an independent distance calibration is supplied.

\appendix

\section{Static and homogeneous limits}\label{app:nonradiative}

The same tensor equations give the static and homogeneous limits of the source projection. For a local normalized bridge with aligned time directions and pressureless sector-$B$ matter,
\begin{equation}
 \PA[T_B]_{00}=\rho_B,
\end{equation}
and the sector-$A$ Newtonian equation is
\begin{equation}
 \nabla_A^2\Phi_A=4\pi G_A
 \left(\rho_A+\epsilon_A\rho_B+\rho_A^{\br}\right).
 \label{eq:Poisson-benchmark}
\end{equation}
The projected density $\epsilon_A\rho_B+\rho_A^{\br}$ is a genuine gravitational source. Its admissible magnitude is restricted by Solar-System dynamics, lensing, cluster offsets, and structure growth~\cite{Will2014,PoissonWill2014,Schneider1992,BartelmannSchneider2001,Clowe2006}.

A matched Friedmann--Lema\^itre--Robertson--Walker (FLRW) slice is defined by
\begin{align}
 t_B=t_A,\quad a_B=a_A\equiv a, \nonumber
 \\ \chi_{AB},\chi_{BA}=\text{constant},
 \label{eq:matched-FLRW}
\end{align}
with pressureless ordinary matter in both sectors. The Friedmann constraints are
\begin{align}
 H^2={}&\frac{8\pi G_A}{3}
 \left(\rho_A+\epsilon_A\chi_{AB}\rho_B+\rho_A^{\br}\right) \nonumber
 \\&+\frac{\Lambda_A}{3}-\frac{k_A}{a^2},\label{eq:Friedmann-A}\\
 H^2={}&\frac{8\pi G_B}{3}
 \left(\rho_B+\epsilon_B\chi_{BA}\rho_A+\rho_B^{\br}\right) \nonumber
 \\&+\frac{\Lambda_B}{3}-\frac{k_B}{a^2}.\label{eq:Friedmann-B}
\end{align}
Their compatibility fixes one bridge density in terms of the other sources,
\begin{align}
 \rho_B^{\br}={}&\frac{G_A}{G_B}
 \left(\rho_A+\epsilon_A\chi_{AB}\rho_B+\rho_A^{\br}\right)
 \nonumber\\
 &-\rho_B-\epsilon_B\chi_{BA}\rho_A
 +\frac{\Lambda_A-\Lambda_B}{8\pi G_B}
 \nonumber\\
 &-\frac{3(k_A-k_B)}{8\pi G_Ba^2}.
 \label{eq:bridge-density-compatibility}
\end{align}
For the dust slice with $\rho_A^{\br}=0$ and separately conserved ordinary and projected dust terms, the spatial bridge stress is fixed by
\begin{equation}
 p_B^{\br}=-\rho_B^{\br}
 -\frac{a}{3}\frac{\dd\rho_B^{\br}}{\dd a}.
 \label{eq:bridge-pressure}
\end{equation}
Equations~\eqref{eq:bridge-density-compatibility} and~\eqref{eq:bridge-pressure} satisfy the homogeneous conservation law in Eq.~\eqref{eq:Bianchi}; together with one Friedmann equation they also supply the corresponding acceleration equation. Thus the matched scale factor is supported by a complete background bridge stress rather than by the Friedmann constraint alone.

\begin{figure}[t]
\centering
\includegraphics[width=0.94\linewidth]{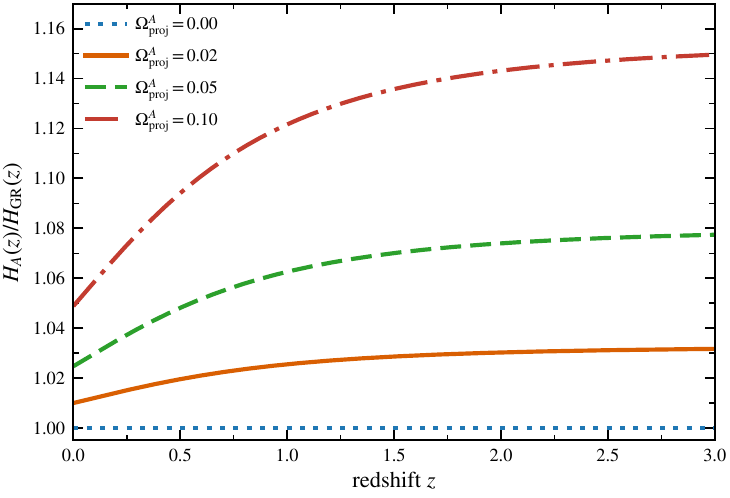}
\caption{Expansion-rate ratio on the matched-dust bridge slice for fixed $\Omega_{m,A}$ and $\Omega_\Lambda$. The projected sector contributes as an additional source.}
\label{fig:cosmology}
\end{figure}

For a flat sector-$A$ illustration with $\rho_A^{\br}=0$, define
\begin{align}
 \Omega_{m,A}&=\frac{8\pi G_A\rho_{A0}}
 {3H_{0,\rm ref}^{2}},\nonumber\\
 \Omega^A_{\proj}&=\frac{8\pi G_A\epsilon_A\chi_{AB}\rho_{B0}}
 {3H_{0,\rm ref}^{2}},\nonumber\\
 \Omega_\Lambda&=\frac{\Lambda_A}{3H_{0,\rm ref}^{2}}.
 \label{eq:Omega-definitions}
\end{align}
At fixed $\Omega_{m,A}$ and $\Omega_\Lambda$,
\begin{align}
 \left[
 \frac{\Omega_{m,A}(1+z)^3+\Omega_\Lambda
 +\Omega^A_{\proj}(1+z)^3}
 {\Omega_{m,A}(1+z)^3+\Omega_\Lambda}
 \right]^{1/2}=\nonumber \\
 \frac{H_A(z)}{H_{\rm GR}(z)}.
 \label{eq:H-ratio}
\end{align}
Figure~\ref{fig:cosmology} shows this direct matterlike contribution~\cite{Weinberg1972,Dodelson2003,Planck2018}. The source-projection coefficient $\Omega^A_{\proj}$ and the radiative angle $\theta$ characterize different bridge sectors.

\backmatter

\section*{Statements and Declarations}

\noindent\textbf{Funding.}
The authors did not receive support from any organization for the submitted work.

\noindent\textbf{Competing interests.}
The authors have no relevant financial or non-financial interests to disclose.

\noindent\textbf{Author contributions.}
Both authors contributed to the conceptual development, methodology, theoretical analysis, interpretation of the results, and revision of the manuscript.

\noindent\textbf{Data availability.}
No new observational data were analyzed in this study. All numerical curves are generated from the analytic equations in the manuscript, and the sensitivity estimate uses published catalog summary constraints.

\noindent\textbf{Use of generative AI.}
OpenAI ChatGPT was used during manuscript preparation for language editing, LaTeX drafting, algebraic cross-checks, and assistance with plotting code. The authors reviewed the equations, references, text, and code and take responsibility for the final manuscript.


\begin{thebibliography}{99}

\bibitem{FierzPauli1939}
Fierz, M., Pauli, W.: On relativistic wave equations for particles of arbitrary spin in an electromagnetic field. Proc. R. Soc. Lond. A \textbf{173}, 211--232 (1939). https://doi.org/10.1098/rspa.1939.0140

\bibitem{BoulwareDeser1972}
Boulware, D.G., Deser, S.: Can gravitation have a finite range? Phys. Rev. D \textbf{6}, 3368--3382 (1972). https://doi.org/10.1103/PhysRevD.6.3368

\bibitem{deRham2011}
de Rham, C., Gabadadze, G., Tolley, A.J.: Resummation of massive gravity. Phys. Rev. Lett. \textbf{106}, 231101 (2011). https://doi.org/10.1103/PhysRevLett.106.231101

\bibitem{HassanRosen2012}
Hassan, S.F., Rosen, R.A.: Bimetric gravity from ghost-free massive gravity. J. High Energy Phys. \textbf{02}, 126 (2012). https://doi.org/10.1007/JHEP02(2012)126

\bibitem{Hinterbichler2012}
Hinterbichler, K.: Theoretical aspects of massive gravity. Rev. Mod. Phys. \textbf{84}, 671--710 (2012). https://doi.org/10.1103/RevModPhys.84.671

\bibitem{deRham2014}
de Rham, C.: Massive gravity. Living Rev. Relativ. \textbf{17}, 7 (2014). https://doi.org/10.12942/lrr-2014-7

\bibitem{SchmidtMay2016}
Schmidt-May, A., von Strauss, M.: Recent developments in bimetric theory. J. Phys. A \textbf{49}, 183001 (2016). https://doi.org/10.1088/1751-8113/49/18/183001

\bibitem{DeFelice2014}
De Felice, A., Nakamura, T., Tanaka, T.: Possible existence of viable models of bi-gravity with detectable graviton oscillations by gravitational wave detectors. Prog. Theor. Exp. Phys. \textbf{2014}, 043E01 (2014). https://doi.org/10.1093/ptep/ptu024

\bibitem{Brizuela2024}
Brizuela, D., de Cesare, M., Soler Oficial, A.: Gravitational wave oscillations in bimetric cosmology. J. Cosmol. Astropart. Phys. \textbf{03}, 004 (2024). https://doi.org/10.1088/1475-7516/2024/03/004

\bibitem{HehlMashhoon2009}
Hehl, F.W., Mashhoon, B.: Nonlocal gravity simulates dark matter. Phys. Rev. D \textbf{79}, 064028 (2009). https://doi.org/10.1103/PhysRevD.79.064028

\bibitem{Milgrom2009}
Milgrom, M.: Bimetric MOND gravity. Phys. Rev. D \textbf{80}, 123536 (2009). https://doi.org/10.1103/PhysRevD.80.123536

\bibitem{Einstein1916}
Einstein, A.: Die Grundlage der allgemeinen Relativit\"atstheorie. Ann. Phys. \textbf{49}, 769--822 (1916)

\bibitem{Hilbert1915}
Hilbert, D.: Die Grundlagen der Physik. Nachr. Ges. Wiss. G\"ottingen Math.-Phys. Kl., 395--407 (1915)

\bibitem{Lovelock1971}
Lovelock, D.: The Einstein tensor and its generalizations. J. Math. Phys. \textbf{12}, 498--501 (1971). https://doi.org/10.1063/1.1665613

\bibitem{Donoghue1994}
Donoghue, J.F.: General relativity as an effective field theory: The leading quantum corrections. Phys. Rev. D \textbf{50}, 3874--3888 (1994). https://doi.org/10.1103/PhysRevD.50.3874

\bibitem{Burgess2004}
Burgess, C.P.: Quantum gravity in everyday life: General relativity as an effective field theory. Living Rev. Relativ. \textbf{7}, 5 (2004). https://doi.org/10.12942/lrr-2004-5

\bibitem{AbishevBerkimbayev2026}
Abishev, M., Berkimbayev, D.Z.: Complementary continuous--discrete time, chronon layering and temporal folding. Symmetry \textbf{18}, 252 (2026). https://doi.org/10.3390/sym18020252

\bibitem{Wald1984}
Wald, R.M.: \textit{General Relativity}. University of Chicago Press, Chicago (1984)

\bibitem{Carroll2004}
Carroll, S.M.: \textit{Spacetime and Geometry: An Introduction to General Relativity}. Addison-Wesley, San Francisco (2004)

\bibitem{Bondi1962}
Bondi, H., van der Burg, M.G.J., Metzner, A.W.K.: Gravitational waves in general relativity. VII. Waves from axi-symmetric isolated systems. Proc. R. Soc. Lond. A \textbf{269}, 21--52 (1962). https://doi.org/10.1098/rspa.1962.0161

\bibitem{Sachs1962}
Sachs, R.K.: Gravitational waves in general relativity. VIII. Waves in asymptotically flat space-time. Proc. R. Soc. Lond. A \textbf{270}, 103--126 (1962). https://doi.org/10.1098/rspa.1962.0206

\bibitem{Maggiore2008}
Maggiore, M.: \textit{Gravitational Waves, Volume 1: Theory and Experiments}. Oxford University Press, Oxford (2008)

\bibitem{Schutz1986}
Schutz, B.F.: Determining the Hubble constant from gravitational wave observations. Nature \textbf{323}, 310--311 (1986). https://doi.org/10.1038/323310a0

\bibitem{HolzHughes2005}
Holz, D.E., Hughes, S.A.: Using gravitational-wave standard sirens. Astrophys. J. \textbf{629}, 15--22 (2005). https://doi.org/10.1086/431341

\bibitem{AbbottGW170817}
Abbott, B.P., et al.: GW170817: Observation of gravitational waves from a binary neutron star inspiral. Phys. Rev. Lett. \textbf{119}, 161101 (2017). https://doi.org/10.1103/PhysRevLett.119.161101

\bibitem{AbbottStandardSiren2017}
Abbott, B.P., et al.: A gravitational-wave standard siren measurement of the Hubble constant. Nature \textbf{551}, 85--88 (2017). https://doi.org/10.1038/nature24471

\bibitem{AbbottGWTC3}
Abbott, R., et al.: Tests of general relativity with GWTC-3. Phys. Rev. D \textbf{112}, 084080 (2025). https://doi.org/10.1103/PhysRevD.112.084080

\bibitem{AbbottCosmologyGWTC3}
Abbott, R., et al.: Constraints on the cosmic expansion history from GWTC-3. Astrophys. J. \textbf{949}, 76 (2023). https://doi.org/10.3847/1538-4357/ac74bb

\bibitem{AbacCosmologyGWTC5}
Abac, A.G., et al.: GWTC-5.0: Constraints on the cosmic expansion rate and modified gravitational-wave propagation. arXiv:2605.27227 (2026). https://doi.org/10.48550/arXiv.2605.27227

\bibitem{Belgacem2018}
Belgacem, E., Dirian, Y., Foffa, S., Maggiore, M.: Gravitational-wave luminosity distance in modified gravity theories. Phys. Rev. D \textbf{97}, 104066 (2018). https://doi.org/10.1103/PhysRevD.97.104066

\bibitem{Will2014}
Will, C.M.: The confrontation between general relativity and experiment. Living Rev. Relativ. \textbf{17}, 4 (2014). https://doi.org/10.12942/lrr-2014-4

\bibitem{PoissonWill2014}
Poisson, E., Will, C.M.: \textit{Gravity: Newtonian, Post-Newtonian, Relativistic}. Cambridge University Press, Cambridge (2014)

\bibitem{Schneider1992}
Schneider, P., Ehlers, J., Falco, E.E.: \textit{Gravitational Lenses}. Springer, Berlin (1992)

\bibitem{BartelmannSchneider2001}
Bartelmann, M., Schneider, P.: Weak gravitational lensing. Phys. Rep. \textbf{340}, 291--472 (2001). https://doi.org/10.1016/S0370-1573(00)00082-X

\bibitem{Clowe2006}
Clowe, D., Brada\v{c}, M., Gonzalez, A.H., Markevitch, M., Randall, S.W., Jones, C., Zaritsky, D.: A direct empirical proof of the existence of dark matter. Astrophys. J. Lett. \textbf{648}, L109--L113 (2006). https://doi.org/10.1086/508162

\bibitem{Weinberg1972}
Weinberg, S.: \textit{Gravitation and Cosmology}. Wiley, New York (1972)

\bibitem{Dodelson2003}
Dodelson, S.: \textit{Modern Cosmology}. Academic Press, San Diego (2003)

\bibitem{Planck2018}
Aghanim, N., et al.: Planck 2018 results. VI. Cosmological parameters. Astron. Astrophys. \textbf{641}, A6 (2020). https://doi.org/10.1051/0004-6361/201833910

\bibitem{Burke1971}
Burke, W.L.: Gravitational radiation damping of slowly moving systems calculated using matched asymptotic expansions. J. Math. Phys. \textbf{12}, 401--418 (1971). https://doi.org/10.1063/1.1665601

\end{thebibliography}
\end{document}